\documentclass[aps,prl,groupedaddress,longbibliography,twocolumn,reprint,amsmath,amssymb]{revtex4-2}
\usepackage{graphicx}
\usepackage{float}
\usepackage{color}
\usepackage{amsmath}
\usepackage{ulem}
\usepackage{bm}
\usepackage{physics}
\usepackage{mathtools}
\usepackage[colorlinks=true,linkcolor=darkblue,anchorcolor=red,citecolor=darkblue, urlcolor=darkblue]{hyperref}
\usepackage[capitalise]{cleveref}
\definecolor{darkblue}{rgb}{0.0, 0.0, 0.55}
\definecolor{light-blue}{rgb}{0.8,0.85,1}
\definecolor{ONANDO}{rgb}{0.047,0.282,0.259}
\definecolor{KURENAI}{rgb}{0.796,0.106,0.270}
\definecolor{PAPER}{rgb}{0.945,0.898,0.788}
\definecolor{apple-green}{rgb}{0.949,0.976,0.929}
\definecolor{MyOrange}{RGB}{254,178,76}

\emergencystretch=\maxdimen
\begin{document}

\title{Transport theory in non-Hermitian systems}

\author{Qing Yan$^{1}$}
\author{Hailong Li$^{1}$}
\author{Qing-Feng Sun$^{1,2}$}\email{sunqf@pku.edu.cn}
\author{X. C. Xie$^{1,2}$}

\affiliation{$^{1}$International Center for Quantum Materials, School of Physics, Peking University, Beijing 100871, China}
\affiliation{$^{2}$CAS Center for Excellence in Topological Quantum Computation,
University of Chinese Academy of Sciences, Beijing 100190, China}

\date{\today}

\begin{abstract}
Non-Hermitian systems have garnered significant attention due to the emergence of novel topology of complex spectra and skin modes. However, investigating transport phenomena in such systems faces obstacles stemming from the non-unitary nature of time evolution. Here, we establish the continuity equation for a general non-Hermitian Hamiltonian in the Schr\"odinger picture. It attributes the universal non-conservativity to the anti-commutation relationship between particle number and non-Hermitian terms. Our work derives a comprehensive current formula for non-Hermitian systems using Green's function, applicable to both time-dependent and steady-state responses. 
To demonstrate the validity of our approach, we calculate the local current in models with one-dimensional and two-dimensional settings, incorporating scattering potentials.
The spatial distribution of local current highlights the widespread non-Hermitian phenomena, including skin modes, non-reciprocal quantum dots, and corner states. Our findings offer valuable insights for advancing theoretical and experimental research in the transport of non-Hermitian systems.

\end{abstract}

\maketitle

\textit{Introduction.---}
Current in physical systems is a response to the external excitation.
Being accurately measured in transport experiments, the current-voltage characteristic and current fluctuations faithfully reflect intrinsic physical properties both statically and dynamically~\cite{Ando1982RMP2Delectronproperty,Kouwenhoven2002RMPdoubleQD,Zwanenburg2013RMPSiquantumelectronics,Stewart1968APLIVofJosephjunc,Science2006CountSingleElectrons}.
Notably, in recently reported topological systems, such as the quantum Hall insulator or the quantum anomalous Hall insulator~\cite{Vonklitzing1980PRLQHE,Changcuizu2013science,zhangyuanbo2020ScienceQAHIMnBiTe}, a precisely quantized current signature shows the robust edge mode protected by the topology of bands ~\cite{HasanKane2010RMPTI,QixiaoliangZSC2011RMPTITSC}.
In these Hermitian quantum systems, the unitary nature of the time-evolution operator ensures the conservation of both the particle number $n$ and current $\bm{j}$, i.e., the continuity equation $\frac{\partial n}{\partial  t}+\mathbf{\nabla } \cdot \bm{j} =0$~\cite{Schrodinger1928CollectedWaveMechanics,Griffiths2018introQM}.

Non-Hermitian systems are in hot spots  for their exotic properties~\cite{Brody2014JPAMTbiorthQM,Ghatak2019JPCMreview,Ashida2020AdvPhysreview,Bergholtz2021RMP}, including non-Hermitian topology, unusual bulk-edge correspondence, skin modes and possible unidirectional amplification~\cite{Fuliang2018PRLtopobandnH,Murakami2018PRLnonBlochband,Lee2016PRLanomalousedge,XiongYe2018JPCbulkedgefailnonHermitian,Kunst2018PRLbiorthBulkBoundary,Zirnstein2021PRLBulkBoundaryGF,Zirnstein2021PRBgrowingbulkGF,wangzhongYao2018PRLnHChernbands,wangzhongYao2018PRLedegetopoinvar,FangchenZhangkai2020PRLwindingandskin,YangZSYi2020PRLnHskinbyonsitediss,Emil2020PRLnHtopoSensor,wangzhong2021PRBLGFnonBloch}. 
For non-Hermitian open chains, skin modes manifest as bulk eigenstates localized at boundaries, exhibiting exponential-decay behavior~\cite{wangzhongYao2018PRLedegetopoinvar}.
Recent progress has focused on the properties of eigenstates which have been observed in both classical and quantum systems 
including optics and photonics, topoelectrical circuits, metamaterials, cold atom systems, and quantum walk systems
~\cite{Ruter2010NatPhysPTinoptics,Weimann2017NatMaterPTphotoniccrystal,Ozdemir2019NatMaterrevEPphotonics,LeeChingHua2018CommunPhys,Helbig2020NatPhysnHcircuits,Ghatak2020PNASnHMechanicalMetamaterial,Lijiaming2019NatCommunPTcoldatom,Xuepeng2020NatPhysnHbulkedge}. 
Beyond their stationary properties, transport techniques can reveal the dynamical response of bulk eigenmodes in non-Hermitian systems. However, exploring the transport properties of these systems remains challenging due to the non-unitary nature of the time evolution operator~\cite{Sticlet2022PRLKubononHermitian}. This non-unitarity further leads to a significant issue during time evolution: the traditional continuity equation requires reevaluation. Given these intricacies, it is crucial to establish the transport theory for non-Hermitian systems.

Addressing the challenges outlined, we investigate the continuity equation for non-Hermitian systems based on the Schr\"odinger picture. We introduce a modified continuity equation, applicable to non-Hermitian Hamiltonians, which incorporates a critical anti-commutation term.  
The anti-commutator indicates a clear distinction from Hermitian systems and directly leads to the phenomenon of non-conservation in non-Hermitian scenarios, affecting physical quantities such as particle number and local current.
Employing Green's function approach, we derive a generalized current formula that captures both temporal and steady-state responses of non-Hermitian systems. Applied to one-dimensional (1D) and two-dimensional (2D) non-Hermitian Hatano-Nelson (HN) models, particularly under the effect of scattering potentials, the current formula reveals the unique features of skin modes, non-reciprocal quantum dots, and corner states. 
Our findings pave the way for further exploration into the dynamic behaviors of non-Hermitian systems.

\textit{Continuity Equation.--}
For any non-Hermitian system, the Hamiltonian can always be decomposed into Hermitian and anti-Hermitian parts, denoted as $\hat{H} = \hat{H}_{\mathrm{H}} + \hat{H}_{\mathrm{A}}$, where $\hat{H}^{\dagger} = \hat{H}_{\mathrm{H}} - \hat{H}_{\mathrm{A}}$. Given a state $\ket{\Phi(t)}$ and the particle number density operator defined as $\hat{n}(\bm{r}) = \hat{\psi}^{\dagger}(\bm{r})\hat{\psi}(\bm{r})$, its expectation value at time $t$ is given by $\langle \hat{n}(\bm{r}) \rangle_t := \bra{\Phi(t)}\hat{n}(\bm{r})\ket{\Phi(t)}$~\cite{Daley2014AdvPhysquantumtrajectory,Barontini2013PRLdissipationColdAtom,Xuepeng2020NatPhysnHbulkedge,descriptionrho}. Provided that the state evolves according to the Schrödinger equation, $i\frac{\partial}{\partial t}\ket{\Phi(t)} = \hat{H}\ket{\Phi(t)}$, then the time evolution of the particle number density is expressed as
\begin{align}
 \! \frac{\partial }{\partial  t}\langle \hat{n} (\bm{r})\rangle_t = \frac{1}{i\hbar}\langle \big[\hat{n} (\bm{r}),\hat{H}_{\mathrm{H}}\big]\rangle_t\! +\! \frac{1}{i\hbar}\langle \big\{ \hat{n} (\bm{r}),\hat{H}_{\mathrm{A}}\big\} \rangle_t ,\label{eq:generalcontinuity}
\end{align}
This represents the general form of the continuity equation in the second quantization form in the Schrödinger picture~\cite{SMnHtrans}. The terms on the r.h.s of \cref{eq:generalcontinuity} correspond to the commutator/anti-commutator between $\hat{n}(\bm{r})$ and the Hermitian/anti-Hermitian parts of the Hamiltonian, $\hat{H}_{\mathrm{H}}$/$\hat{H}_{\mathrm{A}}$, respectively.


We consider the Hamiltionian in the form of $\hat{H}=\hat{H}_0+\hat{H}_{\mathrm{int}}+\hat{H}_{\mathrm{A}}$ 
where the Hermitian part is divided  into non-interacting ($\hat{H}_{0}$) and interacting ($\hat{H}_{\mathrm{int}}$) parts. 
Let $\hat{H}_{0}=\int d\bm{r}\ \hat{\psi}^{\dagger}(\bm{r})a\hat{\bm{p}}^2\hat{\psi}(\bm{r})+\hat{V}(\bm{r})$ represent the non-interacting part. For the anti-Hermitian part, two common non-Hermitian ingredients are considered: $\hat{H}_{\mathrm{A}}=\int d\bm{r}\ \hat{\psi}^{\dagger}(\bm{r})i\bm{\beta}\cdot\hat{\bm{p}}\hat{\psi}(\bm{r})+i\gamma$.
By inserting $\hat{H}$ into \cref{eq:generalcontinuity}, the continuity equation becomes
\begin{align}
  \frac{\partial }{\partial t}\langle \hat{n}(\bm{r} ) \rangle_t +\nabla\cdot \langle \hat{\bm{j}}(\bm{r} ) \rangle_t= \frac{2\gamma}{\hbar}\langle \hat{n}(\bm{r} ) \rangle_t +\frac{\bm{\beta}}{a}\cdot \langle \hat{\bm{j}}(\bm{r} ) \rangle_t,\label{eq:continuiteqHN}
\end{align}
where the local current operator is directly obtained from the first term of $\big[\hat{n}(\bm{r}),\hat{H}_{0}\big]$ in \cref{eq:generalcontinuity}:
\begin{align}
  \hat{\bm{j}}(\bm{r} )=\frac{a}{i \hbar}\left( \hat{\psi}^{\dagger}( \bm{r} )\nabla \hat{\psi}(\bm{r}  )-\left(\nabla \hat{\psi}^{\dagger}( \bm{r}  )\right)\hat{\psi}(\bm{r} ) \right). \label{eq:jop}
\end{align}

Remarkably, both non-Hermitian $\beta$ and $\gamma$ impact and modify the continuity equation \cref{eq:continuiteqHN}, ultimately leading to exotic non-Hermitian phenomena. Unlike non-Hermitian spectra or eigenstates, which are sensitive to boundary conditions, it is essential to emphasize that the continuity equation remains robust regardless of boundary conditions~\cite{SMnHtrans}. 

The continuity equation \cref{eq:continuiteqHN} can accommodate interacting terms, such as the Coulomb interaction 
$\hat{H}_{\mathrm{int}}=\frac{1}{2}\int d\bm{r}d\bm{r}^{\prime}  v (\bm{r},\bm{r}^{\prime} )\hat{\psi}^{\dagger}(\bm{r})\hat{\psi}^{\dagger}(\bm{r}^{\prime})\hat{\psi}(\bm{r}^{\prime})\hat{\psi}(\bm{r})$.
Since the commutator $\big[\hat{n}(\bm{r}),\hat{H}_{\mathrm{int}}\!\big]$ vanishes, these interactions maintain both the form of the continuity equation and the definition of the local current even in the presence of a non-zero $H_A$. 
When the non-Hermitian terms are absent, the continuity equation in~\cref{eq:continuiteqHN} reduces to its Hermitian form, $\frac{\partial}{\partial t}\langle \hat{n}(\bm{r})\rangle_t +\nabla\cdot\langle \hat{\bm{j}}(\bm{r})\rangle_t=0$.

The unique manner of non-conservation of current as indicated by \cref{eq:generalcontinuity} is specific to non-Hermitian systems. While the non-conservative current also exists in Hermitian systems, its underlying cause significantly differs from the scenario presented here. In a Hermitian system, the non-conservation term arises from the non-zero commutator between the particle number operator and the Hamiltonian, for instance, the non-conservative phonons current~\cite{sunqf2007PRBheatgenerationphonons,wangJS2008EPJBquantumthermaltrans}. But for non-Hermitian systems, the correct form of the continuity equation requires the incorporation of anti-commutation relationships, like the two terms on r.h.s of \cref{eq:continuiteqHN} under consideration. Our approach specifically targets the unique correction protocol for the continuity equation when non-Hermitian terms are present, which distinguishes it from the established Hermitian paradigm. 
Up to this point, the phenomena of non-conservation in both Hermitian and non-Hermitian systems have been integrated into a unified framework, as presented in~\cref{eq:generalcontinuity}.

\begin{figure}[t]
  \vspace{ -2pt}  
  \includegraphics[width=0.9\columnwidth]{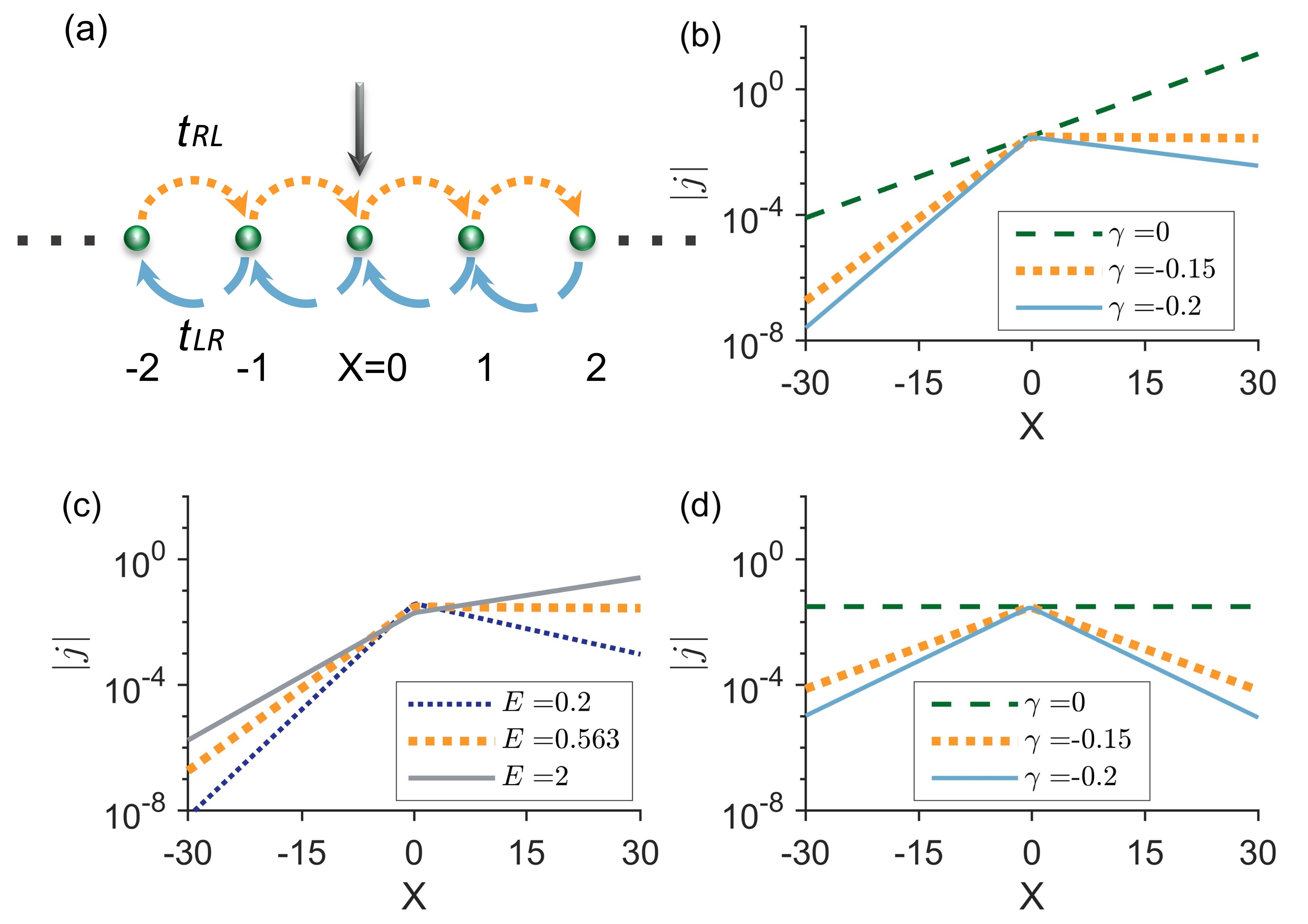}
  \setlength{\abovecaptionskip}{-3pt}   
  \setlength{\belowcaptionskip}{-23pt}   
  \caption{(a) Schematic diagram of an infinite HN chain. The nearest hopping $t_\mathrm{LR}$ and $t_{\mathrm{RL}}$ being discretized from the non-reciprocal term
  ~\cite{SMnHtrans}.
  (b-d) Steady-state local current $j$ of an infinite HN chain with a continuous injection at position $X=0$. The dashed line with $\gamma=0$ in (b) presents the exponentially growing $|j|$ of skin modes. The non-reciprocality in (b) and (c) signifies the existence of skin modes as compared with (d). Parameters are the non-reciprocal term $\beta=0.2$ for (b) and (c), $\beta=0$ for (d), the injecting energy $E=0.563$ in (b) and (d), the on-site loss term $\gamma=-0.15$ in (c). 
  \label{nFig.1}}
\end{figure}

\textit{Green's function and current formula.--}
Green's function is an extremely powerful technique in transport theory, especially for calculating the steady-state and time-dependent current regardless of the system's geometric shape and even in the presence of interactions~\cite{Caroli1971JPCcurrent,Caroli1971JPCcurrentii,MeirWingreenPRL1992Landauer,JauhoWingreenMeirPRB1994,HaugJauho2008quantumtransGF}. For non-Hermitian mesoscopic systems, we are to derive a general current formula expressed by Green's function in both the continuous and discrete forms.

For an arbitrary wavefunction $|\Phi(t)\rangle$, 
the local current $\langle \hat{\bm{j}} \rangle_t$ is calculated as the expectation value of the operator derived from \cref{eq:generalcontinuity}.
Subsequently, we turn to the transport phenomena where the wave function $|\Phi(t)\rangle$ is a response to external excitations, $|\Phi_{0}(t^{\prime})\rangle$. 
Given an arbitrary Hamiltonian $H=H_0+H_{\mathrm{A}}+H_{\mathrm{int}}$, the time-dependent Green's function is defined as~\cite{Economou2006GFinQuantumPhys}
\begin{align}
\left(i\frac{\partial}{\partial t}-H\right)G\left(\bm{r} ,\bm{r}^{\prime};t,t^\prime\right)=\delta\left(\bm{r} -\bm{r}^{\prime}\right)\delta\left(t-t^\prime\right), \label{eq:defGF}
\end{align} 
where $G\!\left(\! \bm{r} \!,\bm{r}^{\prime}\!;\!t ,\!t^\prime\!\right)$  describes how the system at $(\bm{r}, t)$ responds to a pulse excitation at $(\bm{r}^{\prime}, t^{\prime})$ .
Whether the Hamiltonian $H$ is time-dependent, one can solve the corresponding $G$ and obtain the evolved state, $|\Phi(t) \rangle\!=\! i\left[G^r(t,t^\prime\!)-G^a(t,t^\prime\!)\right]|\Phi_{0} (t^\prime)\rangle$. The retarded(advanced) function $G^{r(a)}$ describes the response after(before) a certain excitation. Then, substitute it into \cref{eq:jop} and compute 
the local current $\bm{j}$ at $(\bm{r}, t)$, which is decomposed as $\bm{j}( \bm{r}\!,t )\!=\!\sum_{\mu=1}^{d} j_{\mu}\bm{e}_{\mu}$ with $d$ the dimensionality of the system and $\bm{e}_{\mu}$ being the unit vector in real space. The temporal local current is 
\begin{align}
  j_{\mu}(\bm{r}\! ;t)=-\textstyle\frac{2a}{\hbar}\!\!
  \textstyle\int\!\!   d\bm{r}^{\prime}\!   d\bm{r}^{\prime \prime}
   & \mathrm{Im}\!  [G^r(\bm{r}\! ,\!\bm{r}^{\prime\prime};t,t^\prime)\Phi_{0}(\bm{r}^{\prime\prime},t^\prime)
        \notag \\
   &       \textstyle\frac{\partial}{\partial{\mu}}\! (\Phi_{0}^{\ast}(\!\bm{r}^{\prime},t^\prime\!) G^a(\!\bm{r}^{\prime}\!,\!\bm{r};t^\prime,t\!)\! ) ].\label{eq:j_GF2}
\end{align}

This fomula represents a main result in this paper.
It offers an exact expression for the local current in terms of Green's functions of the non-Hermitian regime.
It remains valid in the presence of various interactions, scattering potentials or random disorder.

We further consider the steady-state transport when the Hamiltionian $H$ is time-independent.
We begin with a continuous and constant injection excited when time $t>0$, characterized by $|\Phi_{0}\rangle$.
Over a sufficiently long period, $t\gg 0$, the system reaches a steady state and the local current is the cumulative response to the injection within the time interval preceding $t$, expressed as $I_{\mu}(\bm{r}) = \int_{-\infty}^t dt^\prime j_{\mu}(\bm{r};t,t^\prime)$.
Within a steady state, local current in \cref{eq:j_GF2} together with Green's functions in \cref{eq:defGF} depend only on the time difference $t-t^{\prime}$ and thus can be Fourier transformed into energy domain,  $I_{\mu}(\bm{r}) = \int_{-\infty}^{+\infty} d\epsilon\ j_{\mu}(\bm{r},\epsilon)$~\cite{SMnHtrans}. Therefore, for a certain injecting energy $\epsilon$, the steady-state local current along the $\mu$-direction at position $\bm{r}$ is

{\setlength{\abovedisplayskip}{2pt}
\setlength{\belowdisplayskip}{-3pt}
\begin{align}
    j_{\mu}(\bm{r},\epsilon)\! = \textstyle\frac{-a}{\pi \hbar}\!
    \textstyle\int\! d\bm{r}^\prime\! d\bm{r}^{\prime \prime} \! & \mathrm{Im}\! \Big[\! G^r(\bm{r}\!,\!\bm{r}^{\prime\prime};\epsilon)\Phi_{0}(\bm{r}^{\prime \prime}\!)  \! \Phi_{0}^{\ast}(\bm{r}^{\prime}\!)\textstyle\frac{\partial}{\partial {\mu}}  G^a(\bm{r}^{\prime}\!,\!\bm{r};\epsilon\!)\!
    \Big] \label{eq:jsteady}
\end{align}}

Here, we discuss the steady-state condition of \cref{eq:jsteady}.
In a Hermitian system, if the Hamiltonian is time-independent, it can always reach a steady state wherein energy and current are conserved~\cite{MeirWingreenPRL1992Landauer}.
However, for non-Hermitian systems, a time-independent Hamiltonian hardly guarantees a steady state, as the presence of non-zero imaginary parts in the energy spectrum obviously disrupts the particle number conservation.
We reveal that achieving a steady state requires that energy spectra possess non-positive imaginary parts; otherwise, achieving a steady state might not be possible. 
Consequently, the derived steady-state current in \cref{eq:jsteady} is applicable to non-Hermitian systems with energy spectra possessing non-positive imaginary components. 

In transport research, the continuous model provides theoretical foundations, while the discrete model offers greater numerical flexibility in modeling realistic non-Hermitian systems. Building on this, we derive the discrete form of the steady-state local current in \cref{eq:jsteady},
{\setlength{\abovedisplayskip}{2pt}
\setlength{\belowdisplayskip}{-3pt}
\begin{align}
  j(\!\bm{R}\!\!+\!\!1\!/\!2\bm{a}_{\mu},\epsilon)\!\!=\!\!\tiny{\frac{-a}{\pi\! \hbar a_0}}\!\!\sum_{\bm{R}^{\prime}\!,\!\bm{R}^{\prime\prime}}\!\! \mathrm{Im}[G^r_{\bm{R},\bm{R}^{\prime\prime}}\!(\epsilon)\Phi_{0}\!(\!{\scriptsize{\bm{R}^{\prime\prime}}}\!)\Phi_{0}^{*}\!(\!\bm{R}^{\prime }\!)G^a_{\bm{R}^{\prime}\!,\!\bm{R}\!+\!\bm{a}_{\mu}}\!\!(\epsilon)].\label{eq:jdiscrete}
\end{align}}
\cref{eq:jdiscrete} represents the local current from site $\bm{R}$ to its neighboring site $\bm{R}\!\!+\!\!\bm{a}_{\mu}$. Here $a_0$ is the discretized lattice constant and $\bm{a}_{\mu}$ is the unit lattice vector in the $\mu-$direction.
{\color{black}Further, for a pluse-like injection, $|\Phi_{0}(\bm{R})|^2=|A|^2/a_0\ \delta_{\bm{R},\bm{R}_{0}}$, at certain position $\bm{R}_0$ with amplitude $A$,
~\cref{eq:jdiscrete} can be simplified as $j(\bm{R}\!+\!1/2\bm{a}_{\mu},\epsilon)\!=\!(1/h) t_{\bm{R},\bm{R}+\bm{a}_{\mu}}^{H}|A|^2  2\mathrm{Im} [G^r_{\bm{R} , \bm{R}_0}\!(\epsilon)G^a_{\bm{R}_0 , \bm{R}+\bm{a}_{\mu}}\!(\epsilon)]$ with  $t_{\bm{R},\bm{R}+\bm{a}_{\mu}}^{H}\!=\! -a/a_0^2$.}
The coefficient  $t_{\bm{R},\bm{R}+\bm{a}_{\mu}}^{H}$ picks the hopping term between neighboring sites in the Hermitian part, $H_0$. 
The non-Hermitian characteristics of the system are captured in the Green's functions, as derived from ~\cref{eq:defGF}, which lead to the emergence of non-conservation in the local current. When $H$ is Hermitian, it reverts to the conventional local current expression~\cite{Jianghua2009PRBtopoAndersInsu,SMnHtrans}.

\textit{Analytical Steady-state current of 1D HN model.--}
Next, we ultilze the derived current formula in~\cref{eq:jsteady} and~\cref{eq:jdiscrete} to analyze the transport properties of some specific non-Hermitian models. 
We first discuss the steady-state current of a 1D HN model of which the Hamiltionian is $\hat{H}=\int dx\ \hat{\psi}^{\dagger}(x)a\hat{p_x}^2\hat{\psi}(x)+\int dx\ \hat{\psi}^{\dagger}(x)i{\beta}\hat{p_x}\hat{\psi}(x)+i\gamma$~\cite{HN1996PRLLocalizationTransitions,HN1997PRBVortexpinning,HN1998PRBnHdelocalization}. Two non-Hermitian terms, $\beta$ and $i\gamma$, correspond to a non-reciprocal hopping term and homogeneous loss term, respectively. 
For an infinite HN chain with a continuous injection,  $|\Phi_{0}(x)|^2=|A|^2\ \delta_{x,x_{0}}$, at certain position $x_0$ with amplitude $A$, we solve the Green's function in~\cref{eq:defGF}, substitute it into~\cref{eq:jsteady}, and thereby the analytical current is derived as~\cite{SMnHtrans},
{
  \setlength{\abovedisplayskip}{1pt}
  \setlength{\belowdisplayskip}{1pt}
\begin{align}
  \!\! j ( x , \epsilon )=\mp\frac{a}{\pi\hbar}|A|^2\frac{\cos{\frac{\varphi}{2}}}{4\sqrt{|z_p|}}e^{\left(\frac{\beta}{a} \pm 2\sqrt{|z_p|}\sin{\frac{\varphi}{2}}\right)(x-x_0)},
  \label{eq:jHN}
\end{align}}
where $z_p\!=\!|z_p|e^{i\varphi}\!=\!(\!\epsilon\!-\!\beta^2/4\!-\!i\gamma)/a$ assuming $\mathrm{Re}z_p\!>\!0$ and $\mathrm{Im}\sqrt{z_p}\!>\!0$. Here, the upper and lower parts in $\mp(\pm)$ correspond to the cases $x\!\! >\!\! x_0$ and $x\!\! <\!\! x_0$.
When the injecting energy exceeds the bottom of the energy spectrum, 
$\epsilon\! >\! \beta^2/4$, a finite current is observed in both left ($x\! <\! x_0$) and right ($x\! >\! x_0$) directions.
For $\beta\!=\!\gamma\!=\!0$, $j$ remains constant on both sides, thereby adhering to Hermitian current conservation. However, when $\beta$ is nonzero, the current exhibits an exponential growth/decay, $j\propto e^{(\beta/a \pm 2\sqrt{|z_p|}\sin{ \varphi/2})x}$. This spatial variation breaks the current conservation but fulfilles the newly derived  continuity equation in~\cref{eq:continuiteqHN}~\footnote{See Section V in the Supplementary Materials~\cite{SMnHtrans}}.

The local current of the HN lattice model in~\cref{nFig.1}(a) is also analytically calculated by~\cref{eq:jdiscrete} with the injection at $x_0=0$~\footnote{Local current in \cref{nFig.1} is analytically calculated by the {G}reen's function based on a discretized lattice Hamiltionian, following the procedure in {Sec. VI} in the {S}upplementary {M}aterials~\cite{SMnHtrans}. {O}ther parameters are $a=1$ and the lattice constant $a_0=0.5$.}.
The local current $|j|$ with $\beta>0$ spatially grows with $\beta$ being the exponential factor in \cref{nFig.1}(b)~(see the dashed line).
Here, the steady-state current with the exponential distribution differs from the constant current of a Hermitian chain in \cref{nFig.1}(d), which matches~\cref{eq:jHN} and also fits the continuity equation~\cite{SMnHtrans}. 
Considering the on-site loss term $\gamma$, the reciprocity of local current distinguishes the non-Hermitian system with/without skin modes, i.e., $|j(-x)|\!\neq\!|j(x)|$ in \cref{nFig.1}(b and c) contrasting $|j(-x)|\!=\!|j(x)|$ in \cref{nFig.1}(d).
With moderate loss, the energy-dependent spatial distribution further highlights such non-reciprocality in \cref{nFig.1}(b). Away from the injection point, $|j(x\!<\!0)|$ consistently decays, while $|j(x\!>\!0)|$ can be decaying, constant or even growing based on injection energy. Thus, the local current characterizes non-Hermitian transport and identifies skin modes through non-reciprocity.

\begin{figure}[t]
  \vspace{-2pt}  
  \includegraphics[width=0.9\columnwidth]{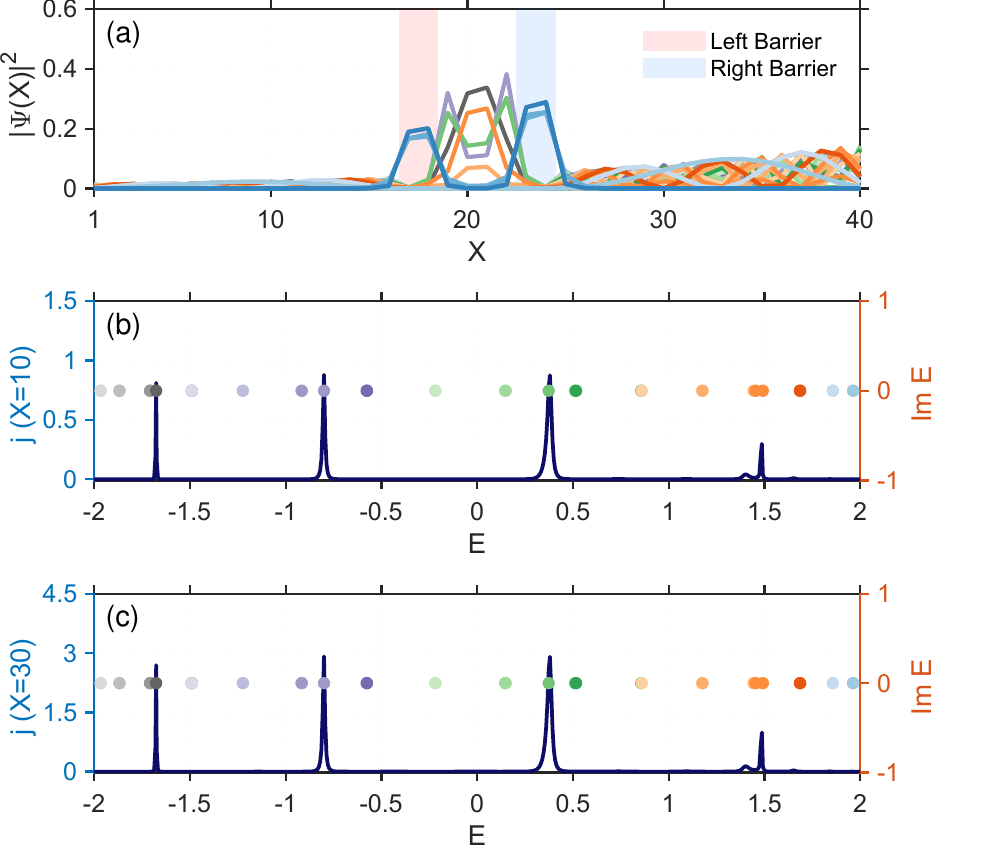}
  \setlength{\abovecaptionskip}{-5pt}   
  \setlength{\belowcaptionskip}{-21pt}   
  \caption{(a)Distribution of eigenstates $|\Psi(\!X\!)|^2$  for a finite HN chain featuring double potential barriers (two shaded regions) reveals an accumulation of skin modes at the right end, while the discrete states formed between the barriers signify non-reciprocal energy levels characteristic of a quantum dot. 
  The corresponding eigenenergies  $(\mathrm{Re}E,\mathrm{Im}E)$ are plotted as colorful scatters in (b and c). The steady-state local current $j$ versus injecting energy $E$ in (b and c) is extracted at sites $X\!=\!10$ and $X\!=\!30$, respectively. The peaks of the local current precisely coincide with the discrete levels of bound states in the quantum dot. 
  Parameters are $\beta=0.05$ and $V_0=3.5$.\label{nFig.2}}
\end{figure}

\textit{Scattering between a double potential barrier.--}
Beyond the homogenous structure, scattering phenomena hold crucial significance within quantum systems. Functional devices can utilize the double barrier setup to form a quantum dot with discrete energy levels accompanied by the resonant tunneling phenomena~\cite{Kouwenhoven2002RMPdoubleQD,Zwanenburg2013RMPSiquantumelectronics,Chang1974resontunneldoublebarrier}.
A symmetrical double barrier is introduced to the 1D non-Hermitian HN model with potential configuration, 
$V\left(x\right)\!=\!V_0\left[ \Theta\left(x-b_{1}-L\right)-\Theta\left(x-b_{1} \right)   \right] \!+\!V_0\left[  \Theta\left(x-b_{2}-L\right)-\Theta\left(x-b_{2}\right)  \right]$
~\footnote{$V_0$ is the height of barriers, $L$ is the length of barriers, and $b_{1}$ and $b_{2}$ mark the left positions of the potential barriers. In the calcualtion of \cref{nFig.2}, the length of HN chain is 40. Other Parameters are $b_{1}=17$, $b_{2}=23$, $L=2$ and $V_0=3.5$  }.
Under open boundary conditions, the eigenfunction in \cref{nFig.2}(a) showcases characteristic skin modes. 
Simultaneously, a bundle of isolated bound states emerges within the spatial constraints of the double barrier, notably with their distribution primarily extending towards the right.

For transport measurements, we couple the system to two terminal leads, enabling electron injection with energy $E$ from the left and calculate the steady-state current by~\cref{eq:jdiscrete}.
Local current emerges only when the incident energy aligns with the discrete levels of bound states in the quantum dot, consistent with quantum resonant tunneling characteristics [\cref{nFig.2}(b and c)].  Interestingly, unlike its Hermitian counterpart, the local current along the non-Hermitian chain does not remain constant in real space, as evidenced by the different magnitude between $j( X = 10 )$ and $j( X = 30 )$. This non-convervation is encoded in the positive non-reciprocal term $\beta$, causing the local current to amplify exponentially from the left injection. Conversely, local current attenuates exponentially from the right injection~\footnote{See the local current result in Section VII in the Supplementary Materials~\cite{SMnHtrans}.}.  As the length of HN chain approaches its thermodynamic limit, unidirectional conductivity only emerges at these resonant tunneling energies. 
This behavior highlights the features of the non-reciprocal quantum dot, confined by double barriers along a non-Hermitian chain: not only does it resonate, but it also selectively amplifies or attenuates signals in a unidirectional manner.

\begin{figure}[t]
  \vspace{-2pt}  
  \includegraphics[width=0.9\columnwidth]{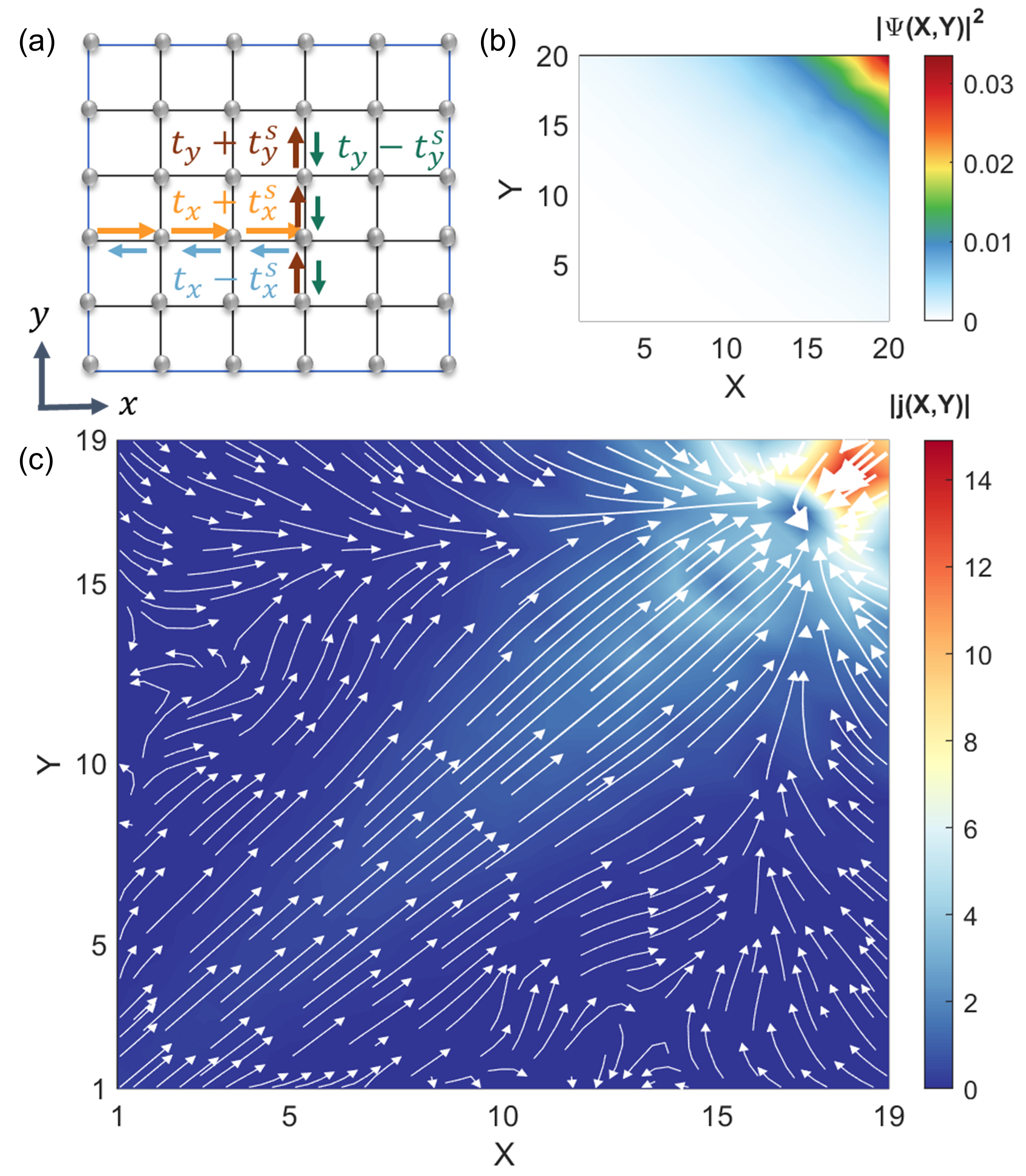}
    \setlength{\abovecaptionskip}{-5pt}   
    \setlength{\belowcaptionskip}{-23pt}   
  \caption{(a) Schematic diagram of a 2D finite HN square with non-reciprocal hopping along the $x$/$y$ direction. 
  (b) Distribution of eigenstates $|\Psi(X,Y)|^2$ of the 2D HN square reveals the corner mode at the top right. 
  (c) A streamline plot depicting the steady-state local current  $|j(X,Y)|$ showcases continuous injection from $(X,Y)\!=\!(1,1)$ and at a certain energy $E\!=\!0.12$. The color reflects the absolute value at each site, while arrows with wavy lines denote the corresponding current direction. Parameters are $t_x=t_y=1, t_x^s=t_y^s=0.1$ and $\gamma=-0.1$.
  \label{nFig.3}}
\end{figure}

\textit{Two-dimensional non-Hermitian regime. --}
The application of the current formula in~\cref{eq:jdiscrete} can also extend to 2D systems. 
In a 2D HN model, non-reciprocal hopping terms are present along both $x$ and $y$ directions, as depicted in~\cref{nFig.3}(a). To investigate the steady-state distribution of local current within the system, we uniformly introduce a dissipation term of $i\gamma$ to each site on the 2D finite square lattice. Under OBC, the eigenstates are localized at the top right [refer to~\cref{nFig.3}(b)], serving as a hallmark of the non-Hermitian corner-skin effect in high-dimensional systems~\cite{zhangkaifangchen2022natcommunUniversalskin}.
Considering the injection of particles with specific energies $E$ from the lower left corner, we proceed to compute the local current of the 2D finite system based on~\cref{eq:jdiscrete}. As illustrated in~\cref{nFig.3}(c), the local current presents with a pronounced increase in magnitude towards the upper right corner. 
This observation distinctly showcases the characteristics of the non-Hermitian corner state.

\textit{Conclusion.---}
Our research introduces a transport theory, focusing specifically on the current response, through the reestablishment of the continuity equation for non-Hermitian Hamiltonians within the Schrödinger picture. We contend that the incorporation of anti-commutators involving non-Hermitian terms plays a fundamental role, leading to a distinctive revision of the continuity equation that sets it apart from its Hermitian counterparts. 
Owing to its universal applicability, this approach allows for a robust examination of the inherent non-conservative current phenomena in non-Hermitian systems.
Employing the Green's function method, we systematically derive an explicit formula for the local current in both temporal and steady-state cases. The universality of this approach is demonstrated across various dimensional examples, unveiling prominent skin modes, non-reciprocal quantum dots, and corner states. Our transport theory establishes a comprehensive theoretical framework, facilitating experimental investigation of non-Hermitian systems.

\textit{Acknowledgement.---}
This work was financially supported by NSF-China (Grant No. 11921005),
the National Key R and D Program of China (Grant No. 2017YFA0303301), the Strategic Priority Research Program of Chinese Academy of Sciences (Grant No. XDB28000000), the National Basic Research Program of China (Grants No. 2015CB921102), the Innovation Program for Quantum Science and Technology (Grants No. 2021ZD0302400), the National Natural Science Foundation of China (Grants No. 12374034 and No. 12304052). Hailong Li is
also funded by China Postdoctoral Science Foundation
(Grant No. BX20220005).


\begin{thebibliography}{60}
\bibitem [{des()}]{descriptionrho}%
    \BibitemOpen
    \href@noop {} {\bibinfo {title} {Although the density of particle number $n$
    may not be conservative, it still serves as a physical observable in
    non-hermitian systems, which can be measured in experiments, such as the
    dissipative cold atom system in~{R}ef.~[39] or the single-photon quantum walk
    system with loss terms in~{R}ef.~[36].}}\BibitemShut {Stop}%
\end{thebibliography}
%
  
\end{document}